\documentclass[useAMS,usenatbib,usegraphicx,epsfig]{aa}
\usepackage{txfonts,graphicx,epsfig}
\def\eps@scaling{.95}
\def\epsscale#1{\gdef\eps@scaling{#1}}
\def\plotone#1{\centering \leavevmode
\epsfxsize=\eps@scaling\columnwidth \epsfbox{#1}}

\def\micron{\mu {\rm m}}

\begin{document}

\title{Monte Carlo Radiative Transfer in  Embedded Prestellar Cores}

\author{D.~Stamatellos\inst{1}\and A.~P.~Whitworth\inst{2}}

\institute{Department of Physics \& Astronomy, Cardiff University, 
        PO Box 913, 5 The Parade, Cardiff CF24 3YB, Wales, UK}
       
   \date{Received March 31, 2003; accepted June 18 , 2003}

\offprints{D.~Stamatellos}

\def\lpacket{$L$-packet }
\def\lpackets{$L$-packets }

\abstract{
We implement a Monte Carlo radiative transfer method, that uses a large
number of monochromatic luminosity packets to represent the radiation 
transported through the system. 
These packets are injected into the system 
and interact stochastically with it. We test our code against various
benchmark calculations and determine the number 
of packets required to obtain accurate results under different circumstances.
We then use this method to study cores that are directly exposed to the
interstellar radiation field (non-embedded cores).
Our code predicts  temperature and
intensity profiles inside these cores which are 
in good agreement with previous studies using  different
radiative transfer methods.

We also explore a large number of models of cores that are embedded in
the centre of a molecular cloud. We study cores 
with different density profiles
embedded in molecular clouds with various optical extinctions
and we calculate temperature profiles, SEDs and intensity profiles.
Our study indicates that the temperature profiles 
in embedded  cores are less steep than those in non-embedded cores. 
Deeply embedded cores (ambient cloud with visual extinction larger 
than 15-25) are almost isothermal at around 7-8 K. The temperature 
inside cores surrounded by an ambient cloud of even moderate thickness 
($A_{\rm V}\sim 5$) is less than 12 K, which is lower than previous 
studies have assumed. Thus, previous mass calculations of embedded cores
(for example in the $\rho$ Ophiuchi protocluster), based
on mm continuum observations,  may underestimate 
core masses by up to a factor of 2.

Our study shows that the best wavelength region to observe embedded
cores is between 400 and 500 $\micron$, where the core is quite distinct from the
background.
We also predict that very sensitive observations ($\sim 1-3$ MJy${\rm\, sr^{-1}}$)
at 170-200 $\micron$ can be used to estimate how deeply a core is
embedded in its parent molecular cloud.  The upcoming {\it Herschel}
mission (ESA, 2007) will, in principle, be able to detect these features 
and test our models. 


}
 
\maketitle

\footnotetext[1]{\email{D.Stamatellos@astro.cf.ac.uk}}
\footnotetext[2]{\email{A.Whitworth@astro.cf.ac.uk}}

\section{Introduction}

The initial stages of star formation are not very well understood. The
general view is that molecular cloud 
cores with sizes $\sim 0.1$ pc and  masses of a few M$_{\sun}$, 
collapse to form stars, either in isolation or in clusters. 
Various observations have been associated with different stages of this
scenario (see Andr\'{e}, Ward-Thompson \& Barsony 2000).
{\it Class 0} 
objects correspond to the first stage in the evolution of a protostar,
where a central luminosity source has been formed in the
centre of the core, and matter accretes on to it.
{\it Class I} objects correspond
to a later stage of collapse where a disc has started to form
around the central object but there is also a residual 
surrounding envelope. Accretion onto the central protostar continues 
but at a lower rate.
{\it Class II} and {\it Class III} objects correspond to the 
classical T Tauri (CTT) and weak-line T Tauri (WTT) stars, 
respectively.
CTT stars have well defined discs, whereas in WTT stars the
inner discs have dissipated.
In addition, observations from Earth (IRAM, SCUBA/JCMT) and space 
({\it IRAS}, {\it ISO})
of various molecular clouds have revealed 
condensations that appear to be on the verge of collapse or already
collapsing
(e.g. Myers and Benson 1983, Ward-Thompson et al. 1994, 
Ward-Thompson et al. 2002, Kirk 2002). 
These condensations are referred to as 
{\it prestellar} cores.

Isolated prestellar cores have been observed inside molecular clouds
(e.g. L1544; Ward-Thompson et al. 1999). These cores are considered 
to be precursors of isolated low mass star formation.
Isolated prestellar cores have extent 
$\stackrel{>}{_\sim} 1.5\times10^4$ AU and  masses
$0.5-35~{\rm M}_{\sun}$ (Andr\'{e}, Ward-Thompson \& Barsony 2000). 
They are not in general
spherically symmetric and they appear to have
flat central density profiles. Magnetic fields are also present
and they may play a role in core stability
(Andr\'{e}, Ward-Thompson \& Barsony 2000 and references therein).

Prestellar cores have also been observed
in young protoclusters, such as $\rho$ Ophiuchi 
(Motte et al. 1998, Johnstone et al. 2000)
and NGC 2068/2071 (Motte et al. 2001, Johnstone et al. 2001).
$\rho$ Oph is a star-forming cluster of about 1~pc diameter,
with estimated average particle density 
$n(H)\sim 2-4\times 10^4\;{\rm cm}^{-3}$
and thermal gas pressure $\sim 10^6\:~{\rm cm}^{-3}$~K
(Liseau et al. 1999).
In this region there have been detected 100 structures, 59 of which
are identified as prestellar cores and the remaining as embedded young
stellar objects (Motte et al. 1998). 
The extent of the prestellar condensations
is  $2-4 \times 10^3$ AU (more compact than isolated prestellar cores), and they 
have
sharp edges. Their estimated masses are  $0.05-3~{\rm M}_{\sun}$.
NGC 2068/2071 are protoclusters in the Orion B cloud complex. Observations 
(Motte et al. 2001, Johnstone et al. 2001) have revealed a filamentary
structure with $\sim 70$  starless condensations having
small sizes ($\sim 5000$ AU) 
and masses from $\sim 0.3~{\rm M}_{\sun}$ to  $\sim 5~{\rm M}_{\sun}$.

Mass estimates from mm continuum observations, where the cores
are optically thin, suggest that the 
initial mass function (IMF) could be determined by 
fragmentation at the pre-stellar stage of star formation 
(e.g.  Andr\'{e}, Ward-Thompson \& Barsony 2000).
The question of whether fragmentation  can produce the
smallest masses in the IMF is still open. Observations of very
low mass prestellar condensations are crucial for answering this
question but they are beyond the limits of today's telescopes. 
Furthermore, current mass estimates are uncertain, 
due to our limited knowledge of the properties of the dust
in and around these cores, and of the dust temperature.
Previous studies have assumed isothermal dust at 12-20 K 
(e.g. Motte et al. 1998, Johnstone et al. 2000). 
More recent radiative transfer studies (Evans et al. 2001, Zucconi et al. 2001)
model cores that are illuminated directly by the isotropic 
interstellar radiation field and 
find that the temperature decreases towards the centre of the core.
However, these studies cannot
be applied to embedded cores, because in this case the illuminating radiation
field is not the interstellar one, and, in general, it is not isotropic.

In this paper, we present a more realistic model that treats 
cores that are  embedded in molecular clouds.
We use a Monte Carlo radiation code we have developed
to study cores approximated by Bonnor-Ebert (BE) spheres. 
In Section~\ref{sec:monte}, we discuss  the basics of our code and 
the tests we have performed to check its validity.
In Section~\ref{sec:system_setup}, we discuss  how
we adapt our code  to treat the radiation transfer in externally
illuminated spheres, and present the tests we have performed.
In Section~\ref{sec:scatter}, 
we briefly examine the effect of different dust properties 
on our calculations and, in Section~\ref{sec:non_embedded}, 
we study BE spheres exposed directly to the Black (1994) 
interstellar radiation field and compare our results, which were
{\it acquired by a different radiative transfer method},
with those of Evans et al. (2001) and Zucconi et al. (2001). 
In Section~\ref{sec:embedded}, we study
the more realistic case of cores embedded in molecular clouds;
we calculate the dust temperature  in their interiors, 
their spectra and their intensity profiles at different 
observing wavelengths. Finally, we summarise our results in 
Section~\ref{sec:conclusions}. 

\section{Monte Carlo Radiative Transfer}
\label{sec:monte}
We implement a method for radiation transfer calculations 
based on a Monte Carlo approach, similar to that developed by
Wolf et al. (1999) and Bjorkman \& Wood (2001).
We make use of the fundamental principle of Monte Carlo methods, according 
to which we can sample a quantity $\xi~\in~[\xi_1,\xi_2]$, from a
probability distribution $p_\xi$ using uniformly distributed random numbers
$\mathcal{R}\in[0,1]$,  by picking $\xi$ such that 
\begin{equation}
\label{mcarlo}
\frac{\int_{\xi_1}^{\xi} p_{\xi^\prime} d{\xi^\prime}} 
{\int_{\xi_1}^{\xi_2} p_{\xi^\prime} d{\xi^\prime}} 
=\mathcal{R} \; .
\end{equation}
Here we briefly outline the basics of our code, named {\sc Phaethon}, 
after a Greek mythical hero.

We discretise the  radiation field of a luminosity source 
(star or background radiation) using a large number
of monochromatic luminosity packets (hereafter referred to as `$L$-packets').
The frequency of an \lpacket is chosen from the source radiation field $I_\nu$, using
Eq.~\ref{mcarlo}, which, in this case, becomes
${\int_0^{\nu_0}{I_\nu(T) d\nu}}/{\int_0^{\infty}{I_\nu(T) d\nu}}=\mathcal{R_\nu}$, 
($\mathcal{R_\nu}\in [0,1]$).
Each of the $L$-packets is injected stochastically into the medium,
 either from a specific point (for a point star)
or from the boundaries of the system (for background radiation). 
For an isotropic radiation field from a point source 
(e.g. Yusef-Zadeh et al. 1984), the direction
$(\theta,\phi)$ of the \lpacket is calculated using 
$\theta=\cos^{-1}(1-2\mathcal{R_\theta})$ and 
$\phi=2\pi\mathcal{R_\phi}\;$ ($\mathcal{R_\theta}\:,\mathcal{R_\phi}\in[0,1]$).
Each \lpacket is also assigned an optical depth, using 
$\tau_\nu=-\ln{\mathcal{R_\tau}}$, $\mathcal{R_\tau}\in [0,1]$, 
and this determines how far the packet propagates into the the 
medium before it interacts with it.

The computational domain in which the \lpackets propagate is divided
into a number of cells.
In regions where the density or the temperature gradients are large, 
more cells are needed. 
We can fulfil both conditions by constructing cells with
dimensions $S_{\rm cell}$
less than, or on the order of, the local directional scale-heights,
\begin{equation}
S_{\rm cell}\stackrel{<}{_\sim}{\rm MIN}\left\{h_\rho,h_T\right\}\; . 
\end{equation}
In the direction given by the unit vector ${\bf k}$, the directional
scale heights are
\begin{equation}
\label{isocells}
h_\rho=
\left(\frac{|{\bf k}\cdot {\bf\nabla}\rho|}{\rho}\right)^{-1}\;\;,\;\;
h_T=\left(\frac{|{\bf k}\cdot  \nabla T|}{T}\right)^{-1}\: .
\end{equation}
In theory, we can construct a grid with a very large number of cells
to satisfy our accuracy requirements in calculating temperature.
However, if we use a large number of cells
we need a large number of \lpackets to interact with each of these cells 
so that the statistical noise of our calculations (of the
order of $1/\sqrt{N_{\rm abs}}$, where $N_{\rm abs}$ is the number of 
\lpackets absorbed in each cell) is small, that will 
increase the computational time of our calculations. 

If $\tau_{\rm total}$ is an $L$-packet's total optical depth then 
in order to calculate
the distance it propagates into the system before it interacts with it, we
need to calculate the line integral along the path of the packet,
\begin{equation}
\Delta S=\int_{0}^{\tau_{total}} \frac{d\tau}{\kappa_\lambda \rho}.
\end{equation}
In the general case
it is not possible to calculate the preceding integral
analytically. 
Our approach is to approximate this integral with a sum:
$\Delta S=\sum_i{({\delta \tau_i}/{\kappa_\lambda \rho_i})}
=\sum_i \delta S_i.$
%
The element step $\delta S_i$ that each \lpacket propagates should be
small, so that the density remains almost constant along this step. Also,
the element optical depth,
$\delta \tau_i=\kappa_\lambda \,\rho_i\, \delta S_i$,
should not be larger than the remaining total optical depth of the \lpacket
$\tau_i$;
$\tau_i$ is just the optical depth that
the \lpacket still has to propagate after $i$ steps,
$\tau_i=\tau_{\rm total}-\sum_{j=0}^{i}\delta\tau_j\;$.
To satisfy the above requirements we chose an element step according to
the following condition:
\begin{equation}
\delta S_i={\rm MIN}
\left\{
{\eta_{\rho} \, h_\rho},
\eta\, l,
(\tau_i+\epsilon)\, l,
\eta_{r}\, |{\bf r}|
\right\},
\end{equation}
where $l=({\kappa_\lambda \rho_i})^{-1}$, and
$\eta_\rho$, $\eta$, $\eta_r$ 
are constants that determine the accuracy we demand
(typical values  are between 0.1 and 1).
The first term (${\eta_{\rho} \, h_\rho}$) ensures that the density does not
change much in one element step ($h_\rho$ is the density scale height
in the direction that the \lpacket propagates), the
second term  ($\eta\, l$) ensures that the element step is less than 
the mean free path of the \lpacket and the third term [$(\tau_i+\epsilon)\, l]$
takes effect on the last step ($\epsilon$ is a very small number).
The last term ($\eta_{r}\, |{\bf r}|$) ensures that the distance the
\lpacket travels in one element step is less than the distance from the 
luminosity
source.
This term comes into effect when a gap exists around the source.
The smaller the factors $\eta_\rho$, $\eta$, $\eta_r$ 
are chosen, the better the
accuracy in propagating \lpackets,  but on the other hand a smaller 
element step means
more computation. 
We propagate the \lpacket following the above procedure until
$
\tau_i \leq 0
$
or  until the packet escapes from the 
system.

When an $L$-packet reaches an interaction point within
the medium (at the end of its optical depth),
it is either scattered or absorbed, depending on the albedo.
If the packet is absorbed its energy is added to the medium and raises
the local temperature. 
To ensure radiative equilibrium the $L$-packet 
is immediately re-emitted.
The new temperature of the cell that absorbs the packet is found by
equating the absorbed and emitted energies.
The re-emitted \lpacket has the same energy but a new frequency
chosen from the difference of the local medium emissivity before and after the
absorption of the packet (Bjorkman \& Wood 2001). The direction
of the reemitted \lpacket is random.
If the \lpacket is scattered then it is assigned a new direction, 
using the scattering phase function due to Henyey \& Greenstein (1941).
Then the packet  propagates again in the medium to a
new interaction location. 
This procedure continues until all the packets escape
from the system. They are then
placed into frequency and direction-of-observation
bins.

\begin{figure}  
\centerline{
\includegraphics[width=7cm,angle=-90]{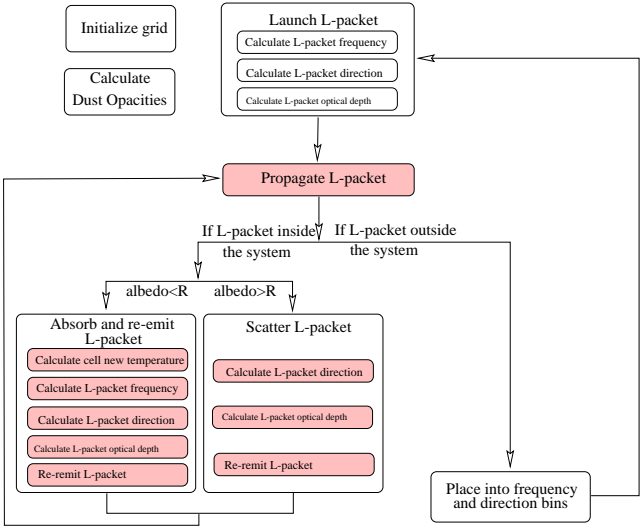}}
\caption{{{\sc phaethon}}: Code flow chart. Luminosity packets are injected into the system,
propagate, interact and finally escape. The propagation routine is the most computationally
expensive routine.}
\label{fig_flowchart_phaethon}
\end{figure}
The general flowchart of the radiative transfer code {\sc Phaethon}
is shown in Fig.~\ref{fig_flowchart_phaethon}.
The shaded parts in this diagram refer to procedures that, in general,
need to be done more than once for each \lpacket and 
consequently are those that dictate the efficiency of the code. 
A simple code efficiency analysis, indicates 
that the {\it \lpacket propagation} routine
takes about 25-50\% of the computational time, depending on the specific
problem. Thus, this is the routine that should be targeted by  any efforts to
diminish the running time of the code.
Time efficiency is very important since a large number of
\lpackets is needed for good statistical results. 
To reduce the running time
of the code, whilst maintaining good results, 
the specific nature of the system we examine should 
be taken into account (for example, in
the case of a uniform density sphere we can propagate each \lpacket in a single
step), any kind of symmetry should be exploited (e.g for spherically
symmetric systems) and look-up tables should be used to solve 
for the cell temperature after absorbing a packet and
then to calculate the reemission frequency of the new packet.

We have tested our code against benchmark calculations proposed by
Ivezic et al. (1997) for a star surrounded by a spherical envelope,
first with constant density, and then for density decreasing as
$r^{-2}$. They used three different, well established 
radiative transfer codes using different numerical schemes to solve 
a set of benchmark spherical geometry problems. In all cases, these methods
gave differences smaller that 0.1\% and, as Ivezic et al. noted, the solution
should be considered  exact. Our code reproduces those results and also
the results of Bjorkman \& Wood (2001), for a disk-like
structure embedded in an envelope.
These tests demonstrate the validity of our radiative transfer code.
We will not present these tests but, instead 
we will later discuss two different tests:
(i) the `thermodynamic equilibrium test' in which
a Bonnor-Ebert sphere is illuminated by a uniform blackbody radiation field
and (ii) the `pure scattering test' in which the albedo of the dust is set
to 1.

This Monte Carlo radiative transfer method conserves energy exactly, accounts for
the diffuse radiation field, can be implemented for any geometrical structure
and is very efficient, making it very attractive for use in a 
variety of problems.
However, it can be implemented, without iteration,  only when the opacity is 
independent of temperature, so the method is useful for 
treating radiation transport against opacity due to
dust grains which are large enough to be in thermal equilibrium.

\section{Radiative Transfer in Prestellar Cores}
\label{sec:system_setup}

\subsection{Core Density Profile}

\begin{figure}  
\centerline{
\includegraphics[angle=-90,width=7.7cm]{3799.f2}}
\caption{Schematic view of Bonnor-Ebert sphere model.\lpackets are ejected from the point
$(0,0,R)$  at such an angle as to imitate an isotropic radiation field.}
\label{fig_sphere}
\end{figure}

A simple approach to  prestellar cores is to assume
that they are isothermal spheres in which gravity is balanced by
gas pressure (Bonnor-Ebert spheres; Bonnor 1956, Ebert 1955).
Recent observations (e.g. Alves et al. 2001, Ward-Thompson et al. 2002) 
show that this is a good approximation for many cores. 
We use the Monte Carlo radiative transfer code to study
cores  
embedded in an isotropic interstellar radiation field.
We choose to parameterise BE spheres using the temperature,  
the mass of the sphere, and the external ambient pressure on the
sphere.
This type of parameterisation is quite useful when examining 
prestellar cores in the same molecular cloud, 
in as much as we can presume that they all experience roughly the same
external pressure. 
The sphere is  divided into a number of concentric cells (typically 50-100) 
with equal radial width (Fig.~\ref{fig_sphere}). 

In our study we assume isothermal gas BE spheres. 
Generally, dust and gas do not have the same temperature unless the density
is quite high, in which case they are thermally coupled 
($n>1-3\times 10^4 \: {\rm cm}^{-3}$, 
Mathis et al. 1983; Whitworth, Francis and Boffin 1998).
However, even non-isothermal models that allow a small gas temperature
gradient, give density profiles that are very close
to the BE profile (Evans et al. 2001). Thus, our results for
isothermal spheres 
should represent non-isothermal spheres reasonably well.

\subsection{The Illuminating Radiation Field}
\label{rad.field}
Because the core is spherically symmetric and the radiation field is isotropic,
we can -without loss of generality- inject all \lpackets at the point 
$(0, 0, R)$ along the $yz$ plane, where $R$ is the radius of the BE
sphere (Fig.~\ref{fig_sphere}). 
If $I_0$ is the integrated intensity of the radiation field,
then the total luminosity incident on the sphere is
$L_{\rm total}=\pi I_0 \: 4 \pi R^2\:.$
If we use $N_p$ luminosity packets, the luminosity per \lpacket is
$\delta L=({\pi I_0 \: 4 \pi R^2})/{N_p}\;$.
For isotropic intensity the injection angle probability is
\begin{equation}
p_\theta d\theta=2\cos{(\theta)} \sin{(\theta)} d\theta\:,\;
 \frac{\pi}{2}\leq\theta\leq\pi\, ,
\end{equation}
and the \lpacket injection angle is therefore 
\begin{equation}
\theta=\cos^{-1}{\left[\mathcal{-R_\theta}^{1/2}\right]},
\;\mathcal{R_\theta}\in [0,1].
\end{equation}

For the spectrum of the radiation incident on the core, we use 
the Black (1994) interstellar radiation field (hereafter BISRF).
Black has compiled an average Galactic background spectrum from
radio frequencies to the Lyman continuum limit, based both on
observations and theoretical modelling (Fig.~\ref{fig_bisrf2}). 
This spectrum consists of an optical component with a peak at around 
1~$\micron$, due to radiation from giant stars and dwarfs; a component due to 
thermal emission from dust grains with a peak at
around 100~$\micron$; mid-infrared radiation from non-thermally heated grains
in the range 5-100~$\micron$; and the cosmic background radiation with a peak
around 1mm ($T=2.728\pm0.004$~K).
This background is similar to that of Mathis et al. (1983) apart from the 
region
from 5 to 400~$\micron$, where it is stronger on the basis of COBE data.
As noted by Black, his estimate only accounts for continuum radiation and
does not include strong emission lines, which may have significant power
in  the far-IR and submillimetre part of the spectrum.

The BISRF seems to be a good approximation to the interstellar
radiation field in the solar neighbourhood. However, it is not always 
an appropriate choice
when studying prestellar cores, because many cores are embedded in molecular
clouds. Consequently, the radiation field is attenuated 
at short
wavelengths ($<~30-40~\micron$) because the surrounding cloud absorbs a large part
of this radiation, and enhanced at long wavelengths ($>50~\micron$) 
due to the thermal emission from the molecular cloud (Mathis et al. 1983).
Also the radiation field may be anisotropic.
In this work, initially we study cores directly exposed to the
BISRF (like the previous studies of Evans et al. 2001 and 
Zucconi et al. 2001)
but we also extend our study to the more realistic case 
of cores inside  molecular clouds of different sizes.

\begin{figure}  
\epsscale{0.7}
\centerline{
\plotone{3799.f3}}
\caption{Black (1994) Interstellar Radiation Field.
}
\label{fig_bisrf2}
\end{figure}

\subsection{Dust Opacities}
\label{sec:dust.opacities}

\begin{figure}  
\epsscale{0.7}
\centerline{
\plotone{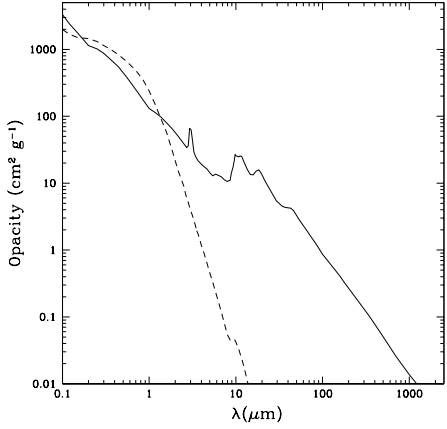}}
\caption{Ossenkopf \& Henning (1994) + MRN (1977) opacities. The solid line represents
the absorption opacity and the dashed line the scattering opacity
($\kappa_{\rm abs}~+~\kappa_{\rm scat}$ $=10.45 \times 10^2\: {\rm cm}^{2}{\rm g^{-1}}$
at $\lambda=0.55~\micron$).
}
\label{fig_opacities}
\end{figure}

Typical dust temperatures in prestellar cores are quite low ($5-20$~K) 
and under
these conditions dust grains are expected to coagulate and
accrete ice mantles.
Following recent studies of prestellar cores (Evans et al. 2001,
Zucconi et al. 2001), we use absorption opacities calculated by Ossenkopf
and Henning (1994) (hereafter OH)
for a standard MRN (Mathis, Rumpl \& Nordsieck 1977) interstellar grain
mixture (53\% silicate \& 47\% graphite), with grains that have coagulated 
and accreted thin ice mantles over a
period of $10^5$ years at densities $10^6 {\rm cm}^{-3}$. 
We also assume a gas-to-dust mass ratio of 100.

Ossenkopf and Henning only calculated absorption opacities 
down to 1~$\micron$, so below this value we use the MRN standard model 
(grains without ice mantles)
with optical constants from Draine \& Lee (1984),
after scaling to fit the OH absorption opacity at
1~$\micron$ (Fig.~\ref{fig_opacities}). 
In any case, the choice of absorption opacities below  1~$\micron$, does not
play an important role in our calculations since at these short wavelengths the
core is optically thick and the radiation does not penetrate much
inside the core.

Also due to lack of data for scattering opacities we use
the MRN scattering opacities after scaling them as before.
We should note though that for dust in  prestellar cores,
scattering is expected to be less by at least a
factor of 2 (Ossenkopf; private communication).
The choice of scattering opacities does not greatly affect 
the temperature in the inner regions of the core, since 
scattering is only important for short wavelength 
($\stackrel{<}{_\sim}$20~$\micron$) radiation, which, anyway, 
cannot penetrate deep inside the core. We will discuss 
the effect of scattering in more detail later.
 
\subsection{Code Tests}

\begin{figure}
\centerline{
\includegraphics[width=4.4cm]{3799.f5}
\includegraphics[width=4.4cm]{3799.f6}}
\caption{Thermodynamic equilibrium Test for a Bonnor-Ebert sphere.
{\bf (a)}~Temperature versus distance from the centre of the sphere.
{\bf (b)}~Spectrum of the incident and the emergent radiation in units of
$B=\sigma T^4/\pi$ (there is no difference at all).
}
\label{fig_be_thermo}
\centerline{
\includegraphics[width=4.3cm]{3799.f7}
\includegraphics[width=4.3cm]{3799.f8}}
\caption{{\bf (a)}~SED of the incident and the emergent radiation 
(no difference at all).
{\bf (b)}~Pure scattering test: Intensity profiles at wavelengths 
850, 200, 450 and 0.55~$\micron$ (top to bottom),
for mean scattering cosine $g=0.5$ . The dotted lines against the righthand
margin of the right plot  correspond to
the background radiation at each of the above wavelengths. 
For $\lambda=450~\micron$
the statistical noise is larger because fewer \lpackets are 
emitted at this wavelength than  at other wavelengths.}
\label{fig_pure_scatter}
\end{figure}

\subsubsection{Test 1: Thermodynamic Equilibrium}
\label{sec:thermo}

Consider a system that is
illuminated by a uniform, isotropic blackbody radiation field
of temperature $T$. Thermodynamic  equilibrium  dictates that
every part of the system
will adopt the same temperature $T$.
This also means that the intensity of the radiation coming from the
system is the same as that of the illuminating blackbody field.    
It is easy to see this from a simple radiative transfer calculation.
If $I_\nu (0)=B_\nu (T)$ is the intensity of the incident radiation
at a specific  direction 
on the system, the intensity $I_\nu (D)$ of the radiation that escapes,
after travelling distance D inside the system, is
\begin{equation}
I_\nu (D)=
I_\nu (0) e^{-\tau_\nu (D)}+ B_\nu (T)  [1-e^{-\tau_\nu (D)}]=B_\nu\:.
\end{equation}
Practically this means that the system is invisible to an observer.
This test can be applied to any 
structure (e.g. spheres, discs, non-symmetric structures) and it is a 
good way to
check the main radiative transfer code (i.e. \lpacket injection, propagation,
absorption, temperature correction and reemission).
It is a very discriminating test and we suggest that 
it should be applied to all radiative transfer codes.

We perform the thermodynamic equilibrium test
for an unstable Bonnor-Ebert sphere 
($\xi_{\rm out}$=11.8, 
$M=4.5~{\rm M}_{\sun}$, $T=11~K$, $P_{\rm ext}=10^4~{\rm cm}^{-3}$~K).
Initially, we do the test  with a blackbody illuminating field having $T=10$~K and
then with $T=20$~K (using $10^9$ luminosity packets).
As seen in Fig.~\ref{fig_be_thermo} the output spectrum 
is the same as that of the illuminating field and the temperature at
any distance from the centre of the core is constant and equal to that
of the radiation field. Small variations on the order of 0.1~K are not
important and are due to statistical noise.

\subsubsection{Test 2: Pure Scattering}

If radiation field incident  on the sphere is isotropic and 
if the \lpackets just pass through the sphere without interacting, the 
observed intensity will be the same at each impact parameter b, and equal 
to the intensity  of the
illuminating field. 
The same holds if the \lpackets just get scattered, i.e. when the albedo
of the grains is set equal to 1. 
It is easy to understand this when the scattering is isotropic, but the same is
also true for non-isotropic scattering. 
Since the incident field is isotropic, 
the \lpackets come from all directions 
and the effect of scattering will simply be to rotate  the whole 
radiation field through an angle $\theta$ but the field will remain isotropic.
The same argument holds if an \lpacket undergoes more than one scattering.
Thus, if the
radiation just gets scattered in the medium,
the emergent spectrum will again be the same as that of the 
incident radiation.

We perform this test for 
a  BE sphere (parameters:
$\xi_{\rm out}=4.1$, $M=4~{\rm M}_{\sun}$, $T=11$~K, 
$P_{\rm ext}=10^4~{\rm cm}^{-3}$~K). 
This time the sphere is illuminated by the BISRF. 
We do calculations for mean scattering cosine
0 (isotropic scattering), 0.5 and 1 (using $2\times10^7$ luminosity packets). 
We present our results for the
$g~=~0.5$ case in Fig.~\ref{fig_pure_scatter}. The code successfully passed
this test too.

\section{Non-Embedded Prestellar Cores}
\label{sec:non_embedded}

We use the term {\it non-embedded} prestellar cores to refer to
cores that are directly exposed to the BISRF.
We perform simulations for a number of 
Bonnor-Ebert spheres under different external pressures and for various 
gas temperatures and masses. 
In Table~\ref{tab:be_runs_params}, we list the parameters 
of our models to show the parameter space investigated. 
BE spheres with the same set of $P_{\rm ext}$,
$T$ and $M$ correspond to one subcritical and one supercritical sphere. 
These can be distinguished by the $\xi_{\rm out}$ value; if $\xi_{\rm out}>6.451$ 
then the sphere is supercritical.
For each model we calculate the temperature profile of the dust 
in the core, the core SED and
intensity profiles at different wavelengths 
(90, 170 and 450 $\micron$).


\begin{table}
\caption{Non-Embedded Prestellar Cores: Model Parameters}
\begin{tabular}{@{}lccclr}
\hline
model ID & $P_{\rm ext}^{\mathrm{a}} ({\rm~K\, cm}^{-3})$  & ${\rm T}^{\mathrm{b}} (~K)$  &  
${\rm M}^{\mathrm{c}} ({\rm M}_{\sun})$  & $\xi_{\rm out}^{\mathrm{d}}$ &$\tau_V^{\mathrm{e}}$ \\
\hline\hline
BE1     & $10^4$  & 10  & 2     &  2.6  &  4.0  \\
BE2     & $10^4$  & 10  & 3.5   &  4.5  &  8.3  \\
BE2.2  & $10^4$  & 10  & 3.5    &  9.9  &  24.0  \\
\hline
BE3     & $10^4$  & 15  & 2     &  1.7  &  2.4  \\
BE4     & $10^4$  & 15  & 4     &  2.4  &  3.6  \\
BE5     & $10^4$  & 15  & 6     &  3.2  &  5.2  \\
BE5.2     &  $10^4$  & 15  & 6  &  21.6  &  64.0  \\
\hline
BE6   & $10^5$  & 15  & 1       &  2.1  &  9.8  \\
BE7     & $10^5$  & 15  & 2.6   &  5.0  &  30.1  \\
BE7.2     & $10^5$  & 15  & 2.6 &  8.5  &  62.0  \\
\hline
BE8     & $5\times 10^4$  & 15  & 2   &  2.6  &  8.9  \\
BE9     & $5\times 10^4$  & 15  & 3.5 &  4.5   &  18.2  \\
BE9.2   & $5\times 10^4$  & 15  & 3.5 &  10.1  &  56.2  \\
\hline
\end{tabular}
\medskip
\begin{list}{}{}
\item[$^{\mathrm{a}}$] External pressure
\item[$^{\mathrm{b}}$] Gas temperature
\item[$^{\mathrm{c}}$] Bonnor-Ebert sphere mass
\item[$^{\mathrm{d}}$] $\xi$ parameter (sphere is supercritical if $\xi>6.451$)
\item[$^{\mathrm{e}}$] Visual optical depth to the centre of the sphere.
\end{list}
\label{tab:be_runs_params}
\end{table}

\begin{figure*} 
\centerline{
\includegraphics[width=5.8cm]{3799.f9}
\includegraphics[width=5.8cm]{3799.f10}
\includegraphics[width=5.8cm]{3799.f11}
}
\caption{Density profiles (a), dust temperature profiles (b) and
SEDs (c), for BE spheres at $T=10$~K under 
external pressure $P_{\rm ext}=10^4 {\rm cm}^{-3}$~K and with masses
3.5 M$_{\sun}$; one subcritical (model BE2, solid lines) 
and one supercritical (model BE2.2, dashed lines). Also 
for a subcritical BE sphere at $T=15$~K with mass 2.6 M$_{\sun}$, under 
external pressure $P_{\rm ext}=10^5~{\rm cm}^{-3}$~K (model BE7, dotted lines).
The dash-dot line on the SED graph corresponds to the background SED.
The temperature at the centre of more centrally condensed cores is lower and 
the core emission is shifted towards longer wavelengths.}
\label{fig_ne1}
\centerline{
\includegraphics[width=5.8cm]{3799.f12}
\includegraphics[width=5.8cm]{3799.f13}
\includegraphics[width=5.8cm]{3799.f14}
}
\caption{Intensity profiles
at 90 (a), 170 (b) and 450~$\micron$ (c), for the models
in Fig.~\ref{fig_ne1} (BE2: solid lines, BE2.2: dashed lines,
BE7: dotted lines).
The horizontal solid lines on the 
profiles correspond to the background intensity at each
wavelength.
At 90~$\micron$ the intensity increases towards the edge of the core but the emission is
just $5-10$~MJy${\rm\, sr^{-1}}$ above the background and, thus,  the cores are barely detectable. 
At 170~$\micron$ the intensity drops towards the edge of the core, or rises by a small amount,
if the core is cold enough.
At 450~$\micron$ the intensity drops towards the edge of the core in all cases.}
\label{fig_ne2}
\end{figure*}

\begin{figure}  
\epsscale{0.9}
\centerline{
\plotone{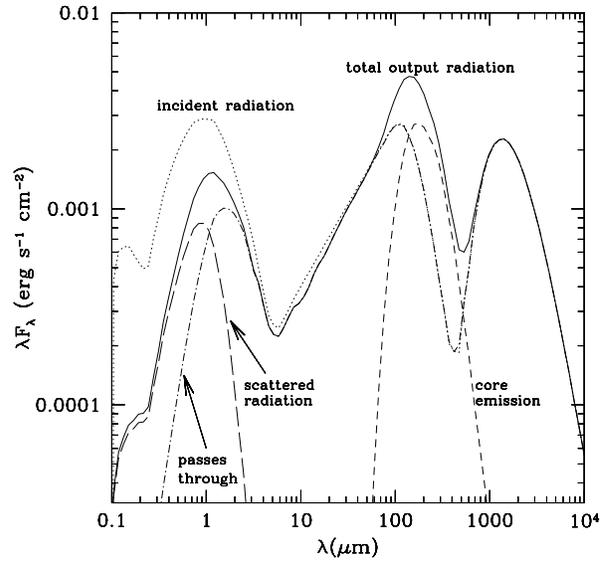}}
\caption{Components of the spectrum for the BE2.2 model.
 The dotted black line is the
SED of the radiation incident on the core, the solid line is the
output SED, the dash-dot line is the part of the radiation that passes
through the core without interacting, the short-dashed line is the 
core emission and the long-dashed line is the scattered light.
}
\label{fig_core}
\end{figure}

\subsection{Temperature Profiles}

The dust temperature inside the core drops from around 17~K at the edge
to a minimum at the centre, which may be as low as
7~K, depending on the total optical depth of
the sphere. 
The higher the optical depth to the centre of the core the lower 
the central temperature and the 
larger the temperature gradient. In Fig.~\ref{fig_ne1}b, 
we plot dust temperature profiles for three representative core models
with different density profiles.

Our results are in general agreement with previous similar calculations
by Evans et al. (2001) and Zucconi et al. (2001).
We compared our results with those of Evans et al. for a system as close we 
could to get to one of their models. 
They do not mention what opacities they use  for $\lambda<1$~$\micron$, 
they use an ISRF at $\lambda<1$, different from the BISRF,
and some parameters in their model are unclear.  
We find that the temperature we calculate at the edge of the core is
$\approx$ 3~K higher than their calculation (17~K rather than 14~K).
This difference can be explained in terms of different opacities and
different ISRF  for $\lambda<1$~$\micron$.
We also find  that the temperature is almost 1~K lower than
reported by Evans et al. at the centre of the 
core. This may in part be
due to slightly different density profiles. 
Zucconi et al. also reported a
higher estimated temperature at the edge ($\approx$ 16.5~K) and a
lower temperature ($\approx$ 0.3~K lower) at the centre of the core, 
when they compared their model with the Evans et al. calculations. However,
these are small differences.

\subsection{SEDs and Intensity Profiles}

We see from the 
component version of the SED (Fig.~\ref{fig_core}), where we plot
the contribution from scattered, processed and direct photons to the SED for
the BE2.2 model, 
that short wavelength radiation ($\lambda\stackrel{<}{_\sim}~$50$~\micron$) 
is absorbed from the core and then is reemitted at longer 
wavelengths, whereas most of the longer wavelength radiation 
($\lambda\stackrel{>}{_\sim}~$50~$\micron$)
just passes through the core without interacting
at all. The UV, optical and NIR radiation that is absorbed is mainly 
responsible for the heating of the core. A large amount of this radiation
will not be available if the core is inside a molecular cloud, as we discuss
later in this paper.
The core emits most of its radiation in the FIR and submm 
(also see Fig.~\ref{fig_ne1}c).
The peak of the emission is between 110 and
160~$\micron$ (note that this is the peak of $\lambda F_\lambda$ not
$F_\lambda$). At these wavelengths, the core 
is easily observable against the background.   
At shorter wavelengths (e.g. 90~$\micron$) the contrast with
the background is not very distinct.
Finally, in the optical the core is seen in absorption and 
appears like a black blob against the bright background.

The radial intensity profile of a core at a specific wavelength  
$\lambda_{\rm obs}$
depends on whether this wavelength  is shorter or longer from the peak
wavelength $\lambda_{\rm peak}$. 
If $\lambda_{\rm obs}$ is much  
longer than $\lambda_{\rm peak}$ (e.g. at 450~$\micron$)
then the Rayleigh-Jeans approximation
for the Planck function holds and the intensity is proportional to the product
of column density with temperature. The density decrease towards the edge of the
core is much larger than the corresponding temperature increase, and  
the intensity decreases considerably towards the edge 
(see Fig.~\ref{fig_ne2}c). 
If  $\lambda_{\rm obs}$ is much shorter than $\lambda_{\rm peak}$ 
(e.g. at 90~$\micron$) 
then the Wien approximation
holds and the intensity depends on the temperature
exponentially,
so even a small increase in the temperature can balance the density decrease
and the intensity increases slightly ($\sim 5$ MJy${\rm\, sr^{-1}}$) towards the edge 
of the core (Fig.~\ref{fig_ne2}a). 
However, the contrast between core and background radiation is very small 
($\sim 5-7$ MJy${\rm\, sr^{-1}}$) and
cores should be barely detectable at 90~$\micron$. This result is consistent
with observations of prestellar cores (Ward-Thompson et al. 2002) that
show that cores are usually well defined at 170 and 200~$\micron$ but not 
always well defined at 90~$\micron$. 
Finally if  $\lambda$ and $\lambda_{\rm peak}$ 
are comparable (e.g. 170~$\micron$)
the intensity either drops from the centre to the edge 
($\lambda$ a bit longer than $\lambda_{\rm peak}$; 
Fig.~\ref{fig_ne2}b, models BE2 and BE2.2) or it
increases ($\lambda$ a bit smaller than $\lambda_{\rm peak}$;
Fig.~\ref{fig_ne2}b, model BE7).
In general, the contrast with the background is quite large at these intermediate
wavelengths.
\begin{figure*} 
\centerline{
\includegraphics[width=5.8cm]{3799.f16}
\includegraphics[width=5.8cm]{3799.f17}
\includegraphics[width=5.8cm]{3799.f18}
}
\caption{Temperature profiles (a) and intensity profiles at 0.55
$\micron$ (b), 170 and 450 $\micron$ (c) for a Bonnor-Ebert sphere 
($\xi_{\rm out}$=11.8, $M=4~{\rm M}_{\sun}$, 
$T=11$~K, $P_{\rm ext}=10^4 {\rm cm}^{-3}$~K) with different dust
scattering properties. The dashed line corresponds
to mean scattering cosine $g=1$ (forward scattering, i.e. in effect, no scattering), the dotted line 
to $g=0.4$ and the solid line to $g=0\;$. The dash-dot horizontal lines on 
the intensity profiles correspond to the background intensity at
the wavelenght noted on the graph. Different dust mean scattering cosines do not
greatly affect the dust temperature profile in the core.}
\label{fig_scatter1}
\centerline{
\includegraphics[width=5.8cm]{3799.f19}
\includegraphics[width=5.8cm]{3799.f20}
\includegraphics[width=5.8cm]{3799.f21}
}
\caption{Temperature profiles (a) and intensity profiles at 0.55
$\micron$ (b), 170 and 450 $\micron$ (c) for a Bonnor-Ebert sphere 
($\xi_{\rm out}$=11.8, $M=4~{\rm M}_{\sun}$, 
$T=11$~K, $P_{\rm ext}=10^4 {\rm cm}^{-3}$~K) with different dust
scattering opacities. The dashed line corresponds to zero scattering
opacity, the dotted line to half the MRN scattering opacity and
the solid line to the MRN standard model scattering opacity.
 The dash-dot horizontal lines on 
the intensity profiles correspond to the background intensity at
the wavelength noted on the graph. Different dust scattering opacities
do not greatly affect the dust temperature profile in the core.}
 \label{fig_scatter2}
\end{figure*}

\subsection{Effects of Dust Scattering Properties}
\label{sec:scatter}

The properties of dust in molecular clouds and 
prestellar cores are quite uncertain (see Andr\'{e}, Ward-Thompson \& Barsony 2000).
In this section we examine the effect of different dust scattering
properties on the temperature profiles and on the spectra of prestellar 
cores. 
We perform radiative transfer calculations using {\sc Phaethon},
for a supercritical Bonnor-Ebert sphere 
($\xi_{\rm out}$=11.8, $M=4~{\rm M}_{\sun}$, 
$T=11$~K, $P_{\rm ext}=10^4 {\rm cm}^{-3}$~K) with
total visual optical depth $\tau_{\rm V}=30.6$, and different dust properties.
 
Initially, we vary the mean scattering cosine $g$. 
We  see (Fig.~\ref{fig_scatter1}a) 
that when the scattering is isotropic ($g=0$, solid line) the dust 
temperature at a specific radius inside the core is a bit lower 
($\sim 0.3~K$) than for the case of forward scattering ($g=1$, dashed line).
The case $g=1$ is equivalent to no scattering, so at optical wavelengths there is
significant intensity only at the very edge of the core where the optical depth
through the core is small and radiation can pass straight through (Fig.~\ref{fig_scatter1}b).
If photons are scattered forward, they are able to penetrate deeper inside
the core and heat it to higher temperatures. 
As a result more optical photons are absorbed
and more FIR photons are emitted.
The intensity difference between dust models with different mean scattering cosine
is very large in the optical region (Fig.~\ref{fig_scatter1}b)
but it is only  $\sim 10-20\%$ at  FIR 
and submillimetre wavelengths (e.g. 170 and 450 $\micron$, Fig.~\ref{fig_scatter1}c).

Next, we vary the scattering opacities of the dust 
(Fig.~\ref{fig_scatter2}).
We perform 3 calculations: 
(a) with MRN scattering opacities 
$\kappa_{\rm scat}=\kappa_{\rm scat}^{\rm MRN}$ (solid lines), 
(b) with $\kappa_{\rm scat}=\kappa_{\rm scat}^{\rm MRN}/2$ (dotted lines), 
and (c) with no scattering at all ($\kappa_{\rm scat}=0$, dashed lines).
The results are similar to the previous case:
when there is no scattering (which is the same as $g=1$ in
Fig.~\ref{fig_scatter1}) more photons are absorbed by the
core, heating it to slightly higher temperatures. 
Scattering provides photons with a quick way out 
of the core without them being absorbed.

This study shows that different dust composition, as 
reflected in different dust scattering opacity and
different scattering mean cosine, results in only slightly different
temperature profiles. The optical intensity profiles are strongly dependent
on the dust scattering properties but at FIR and submillimetre wavelengths,
where prestellar cores emit most of their radiation, the intensity is not affected significantly. 
Thus, we conclude that the scattering properties of the dust do not
greatly affect the results of our radiative transfer calculations of prestellar cores.

\section{Prestellar Cores Embedded in Molecular Clouds}
\label{sec:embedded}

\begin{figure}  
\epsscale{0.7}
\centerline{
\plotone{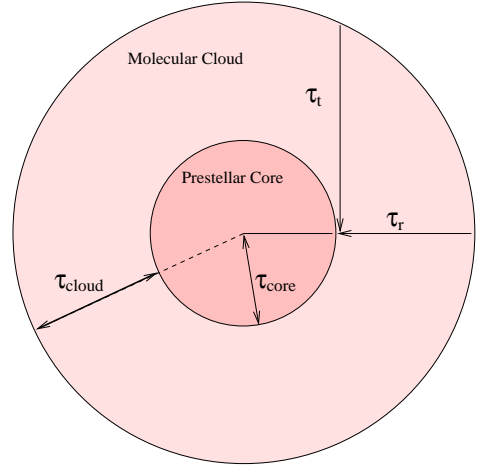}}
\caption{Schematic representation of a prestellar core embedded in a 
molecular cloud (not in scale). The radiation incident on the core is not
isotropic because $\tau_{\rm t}>\tau_{\rm r}=\tau_{\rm cloud}$.}
\label{fig_core_cloud}
\end{figure}

In many cases prestellar cores are embedded deep inside molecular clouds
and the radiation incident on them is 
different from the interstellar radiation field, 
and anisotropic (Mathis et al. 1983).
The ambient molecular cloud acts like a shield to
UV, visual and NIR interstellar radiation, absorbing 
and re-emitting it in the FIR.    
It also makes the radiation incident on the core anisotropic 
because in general the molecular cloud is not  homogeneous
and it is not spherically symmetric. 
Even for a spherical  ambient molecular cloud with uniform density 
the radiation incident on the prestellar core is not isotropic,
even if the core lies at the centre of the molecular cloud. 
That is because there will be more
radiation incident on a specific point on the embedded core
from the radial direction (which is closer to
the boundary of the cloud) than from the tangent or any other direction 
(see Fig.~\ref{fig_core_cloud}).   

Another factor  contributing to the anisotropy of
the radiation incident on a prestellar
core is the presence of stars or other luminosity sources in the vicinity of 
the core.
For example, according to the Liseau et al. (1999) model there is
a B2V star close to $\rho$ Ophiuchi that increases the UV radiation
incident on the cloud from one side.
Also the NGC 2068/2071 protoclusters in Orion B (Motte et al. 2001) are in an
environment rich in FIR, submm and mm photons, from reprocessed UV radiation 
from the
newly born stars in Orion. In such cases, the BISRF is 
probably not a very good representation of the radiation field incident on the core.

Previous studies (Evans et al. 2001, Young et al. 2002) have acknowledged
that deviations from the BISRF are important and 
have used a scaled version of the BISRF that 
is either enhanced at all wavelengths
or selectively at UV and FIR. This simple approach has a free parameter, 
the ISRF scaling factor, that is varied arbitrarily to fit the observations but
it is not connected directly to the molecular cloud in which the core is
 embedded
or the transport of radiation inside the cloud, 
and does not account for the fact that the radiation field incident on 
an embedded core is not isotropic.
Here, we present more consistent models in which deviation from the
BISRF is a direct result of the presence of a molecular cloud that
surrounds the core. 

\subsection{Model Description}


\begin{table*}[ht]
\caption{Embedded Prestellar Cores: Model Parameters}
\begin{center}
\begin{tabular}{lccccccccc}
\hline
model ID      & $M^{\mathrm{a}} ({\rm M}_{\sun})$  & $T^{\mathrm{a}}$ (K)   & ${\xi_{\rm out}}^{\mathrm{a}}$   & ${n_{\rm c}}^{\mathrm{a}} ({\rm cm}^{-3})$ 
& ${n_{\rm b}}^{\mathrm{a}} ({\rm cm}^{-3})$ & ${\tau_V}^{\mathrm{a}}$  & ${R_{\rm BE}}^{\mathrm{a}}$ (AU) & ${\tau_{\rm cloud}}^{\mathrm{b}}$ 
& ${R_{\rm cloud}}^{\mathrm{b}}$ (AU) \\
\hline\hline
EM1        &  & & & & & &    &  0   &  6.1 $\times 10^3$    \\
EM1.05     &   0.8  & 15   & 4.7     & 4.5 $\times 10^5$  & 6.7 $\times 10^4$  & 87  & 6.1 $\times 10^3$&  5   &  1.5 $\times 10^4$   \\
EM1.10     &   & & & & & &  &  10  &  2.3 $\times 10^4$   \\
EM1.20     &   & & & & & & &  20   & 4.1 $\times 10^4$   \\
\hline
EM2        &   & & & & & &  &  0   &  5.2 $\times 10^3$   \\
EM2.05     & 0.8  & 15   & 9.4     & 2.4 $\times 10^5$   & 6.7 $\times 10^4$ & 227  & 5.2 $\times 10^3$    &  5   &  1.4 $\times 10^4$     \\
EM2.10     &   & & & & & & &  10   & 2.3 $\times 10^4$     \\
EM2.20     &  & & & & & & &  20   & 4.0 $\times 10^4$     \\
\hline
EM3        &  & & & & & & &  0  &   5.4  $\times 10^3$   \\
EM3.05     & 0.4  & 15   &  2.4    & 1.4 $\times 10^5$   & 6.7 $\times 10^4$&  36   &  5.4  $\times 10^3$    &  5  &  1.4 $\times 10^4$      \\
EM3.10     & & & & & & & &  10  &   2.3 $\times 10^4$    \\
EM3.20     & & & & & & &  &  20  &  4.0 $\times 10^4$      \\
\hline
EM4        & & & & & & & &    0     & 8.1 $\times 10^3$     \\
EM4.05     & 1.2  & 20   &  3.6    & 1.9 $\times 10^5$  &  5.0 $\times 10^4$  &  61  &  8.1  $\times 10^3$   &  5       &  2.0 $\times 10^4$    \\
EM4.10     & & & & & & &  &  10       & 3.1 $\times 10^4$    \\
EM4.20     & & & & & & &  &  20       &  5.5 $\times 10^4$    \\
\hline
EM5        & & & & & & & &  0        & 6.2 $\times 10^3$     \\
EM5.05     & 1.2  &  20  &  15.4   &  6.1 $\times 10^6$ &  5.0 $\times 10^4$   & 430   &   6.2  $\times 10^3$ &  5        & 1.8 $\times 10^4$     \\
EM5.10     & & & & & & &  & 10        & 3.0 $\times 10^4$     \\
EM5.20     & & & & & & &  & 20        & 5.3 $\times 10^4$     \\
\hline
\end{tabular}
\end{center}
\begin{list}{}{}
\item[$^{\mathrm{a}}$] Embedded core properties: $M$: mass, $T$: gas temperature, ${\xi_{\rm out}}$: $\xi$ parameter
of the BE sphere ($\xi>6.451$ for supercritical spheres), ${n_{\rm c}}$: central density,
${n_{\rm b}}$: boundary density, ${\tau_V}$: visual optical depth to the centre of
the sphere, ${R_{\rm BE}}$: radius of the sphere.
\item[$^{\mathrm{b}}$] Ambient cloud properties: ${\tau_{\rm cloud}}$: visual optical 
the of the cloud (see Fig.~\ref{fig_core_cloud}), ${R_{\rm cloud}}$: cloud radius.
\end{list}
\label{tab:em_runs}
\end{table*}

We examine the simple model of a spherical prestellar core 
that is at the centre of a 
spherical molecular cloud (see Fig.~\ref{fig_core_cloud}). 
We try to mimic the conditions in $\rho$ Ophiuchi, where 
condensations have masses in the range 0.05-3 ${\rm M}_{\sun}$ and dimensions
$1-6 \times 10^3$ AU ($\sim 7-42$ arcsec). The thermal pressure at the edge of
the cloud is $\sim 10^6$ cm$^{-3}$~K  and the 
estimated particle density is $\sim 2\times 10^4\;{\rm cm}^{-3}$
(Liseau et al. 1999). 
In our study we examine cores with dimensions
$4-8 \times 10^3$ AU and masses $0.4-1.2~{\rm M}_{\sun}$.
We assume that the molecular
cloud outside the core 
has constant particle density $n(H_2)= 0.77\times 10^4\;{\rm cm}^{-3}$ 
(corresponding to  $n_{\rm tot}=0.96 \times 10^4\;{\rm cm}^{-3}$ 
for a gas with mean molecular
weight $\mu=2.3$ and hydrogen abundance by mass  $X=0.7$).  
We also assume that 
the dust in the molecular cloud has the same  composition as the dust in 
the core and, therefore, the same opacities. As in the study of
non-embedded cores we use the Ossenkopf and Henning (1994) opacities  (see 
Section~\ref{sec:dust.opacities}).
We use a BE sphere density profile for the prestellar cores,
with fixed 
ambient pressure at $\sim 10^6\; $cm$^{-3}$~K.
Thus, the free parameters
in defining the BE profile are the mass of the sphere and the gas 
temperature. 
Because $n(H_2)$ is specified,
the visual optical depth $\tau_{\rm cloud}$ is the only free parameter 
for the ambient cloud. $\tau_{\rm cloud}$  also determines the extent 
of the cloud. 
Motte et al. (1998)
calculate $A_V\sim10$ mag for $\rho$ Ophiuchi, but depending on the position
of the core in the cloud, 
the extinction could be up to $\sim 40$ mag. We use visual optical depths 5,
10 and 20.
The detailed parameters of our models are listed in 
Table~\ref{tab:em_runs}.

\subsection{Temperature Profiles and Mass Estimates}

The dust temperature profile 
inside the  core depends on the optical depth
 of the molecular cloud
in which the core is embedded, 
and the density profile of the core. (Additionally, 
the dust opacities are important, but we will not study their
influence here.)
The presence of even a moderately thick cloud 
of $\tau_{\rm cloud}=5$ around the core, 
shields the core from
UV and NIR radiation, resulting in a less steep 
temperature profile inside the core than in the case of a core
that is directly exposed to the interstellar radiation field.
When there is no surrounding cloud the temperature drops
from 16~K at the edge of the core to around 6-7~K in the centre 
($\Delta T\approx 9-10$~K, 
depending on the core density), whereas with a   
$\tau_V=5$ cloud the temperature drops from  around 11~K to 7~K 
($\Delta T\approx 4$~K), as seen in
Figs.~\ref{fig_oem1a}b and \ref{fig_oem2a}b. 
Particularly, in the case of a deeply embedded core ($\tau_{\rm cloud}=20$)
the core is almost isothermal ($\Delta T\stackrel{<}{_\sim}1.5$~K) at around
7-8~K, for a not very centrally condensed core 
(Fig.~\ref{fig_oem1a}b), whereas for a supercritical
core (Fig.~\ref{fig_oem2a}b) $\Delta T\approx 3$~K. 
Our studies show that temperatures inside embedded cores are
probably lower than 12~K
in cores surrounded by even a relatively thin  cloud
 (visual $\tau_{\rm cloud}=5$),
which seems to
be the case for many of the prestellar cores in $\rho$ Ophiuchi.
Previous studies (Motte et al. 1998, Johnstone et al. 2000) 
of prestellar cores in $\rho$ Oph
assumed isothermal dust at temperatures from 12 to 20~K, 
when calculating core masses from mm observations. 
At these wavelengths the dust emission is optically thin
and, consequently, the observed flux is 
\begin{equation}
F_{\rm \lambda}=B_{\rm \lambda}(T_{\rm dust})\:\tau_{\rm \lambda}\:\Delta\Omega
=B_{\rm \lambda}(T_{\rm dust})\:\kappa_{\rm \lambda}\,{N(H_2)}\,{\mu m_H}\:
\Delta\Omega.
\end{equation}
Hence, the inferred column density is 
\begin{equation}
N(H_2)=\frac{F_{\rm \lambda}}{\mu m_{H}\:\Delta\Omega\; \kappa_{\rm \lambda}\: 
B_{\rm \lambda} (T_{\rm dust})}\;.
\end{equation}
$\Delta\Omega$ is the solid angle of the telescope beam for a resolved source,
or the solid angle of the source if unresolved,
$N(H_2)$ is the
hydrogen column density,
$\kappa_{\rm \lambda}$ is the mm dust opacity, and B$_{\rm \lambda}$ is the
Planck function.
At mm wavelengths and temperatures $<20~K$ the Rayleigh-Jeans approximation holds, so
$B_{\rm \lambda}(T_{\rm dust})\propto T_{\rm dust}$. Therefore,
the estimated column density, and consequently the mass, depends on 
the observed mm flux, and the dust opacity and temperature,
\begin{equation}
N(H_2)\propto \frac{F_{\rm \lambda}}{\kappa_{\rm \lambda}\:T_{\rm dust}}\;.
\end{equation}
Thus, the masses of the prestellar condensations calculated by Motte et al.
and Johnstone et al., using mm continuum observations,  may be 
underestimated by up to a factor of 2, which will affect their 
evaluation of the core mass function in 
the $\rho$ Oph protocluster, and the 
inferred stability or instability of the observed cores.
Detailed modelling for each of the prestellar
cores, taking into account their environment (i.e. surrounding cloud
and nearby luminosity sources), is needed to
calculate their masses with more accuracy. Also, as Motte et al. point out,
the dust opacities and also the dust-to-gas ratio, 
introduce additional uncertainties in mass calculations.

\begin{figure*} 
\centerline{
\includegraphics[width=5.8cm]{3799.f23}
\includegraphics[width=5.8cm]{3799.f24}
\includegraphics[width=5.8cm]{3799.f25}
}
\caption{Density profiles (a), dust temperature profiles (b) and
SEDs (c), for BE spheres at $T$=15~K with mass 0.8 M$_{\sun}$,
under external pressure $P_{\rm ext}=10^6 {\rm cm}^{-3}$~K, surrounded
by a spherical ambient cloud with visual optical depth 20 (model EM1.20,
solid lines), 5 (model EM1.05, dotted lines) and 0 (model EM1, dashed lines;
no surrounding cloud).
The dash-dot line on the SED graph corresponds to the background SED.
The deeper the core is embedded the lower the dust temperature inside the core.
The dust temperature is lower than 12~K even when the core is embedded in 
a relatively thin molecular cloud with visual extinction 5 mag.}
\label{fig_oem1a}
\centerline{
\includegraphics[width=5.8cm]{3799.f26}
\includegraphics[width=5.8cm]{3799.f27}
\includegraphics[width=5.8cm]{3799.f28}
}
\caption{Intensity profiles
at 90 (a), 170 (b), 450 and 850~$\micron$ (c), for the models
in Fig.~\ref{fig_oem1a} (EM1.20: solid lines, EM1.05: dotted lines, 
EM1: dashed lines).
The horizontal solid lines on the 
profiles correspond to the background intensity at the
wavelength marked on the graph.
At 90~$\micron$ the core is seen in absorption against the background but 
the core is not easily distinguishable.
At 170~$\micron$ the intensity increases towards the edge of the core
only if the core is not very deeply embedded. However, very
sensitive observations are needed to detect this feature.
At 450 and 850~$\micron$ the intensity drops towards the edge of the core.}
\label{fig_oem1b}
\end{figure*}

\subsection{SEDs and Intensity Profiles}

\begin{figure*} 
\centerline{
\includegraphics[width=5.8cm]{3799.f29}
\includegraphics[width=5.8cm]{3799.f30}
\includegraphics[width=5.8cm]{3799.f31}
}
\caption{The same as in Fig.~\ref{fig_oem1a}, but for a supercritical BE
sphere with the same parameters (EM2x models): 
Density profiles (a), dust temperature profiles (b) and
SEDs (c), for BE spheres at 15~K, with mass 0.8 M$_{\sun}$,
under external pressure P$_{\rm ext}=10^6 {\rm cm}^{-3}$~K, surrounded
by a spherical ambient cloud with visual optical depth 20 (model EM2.20,
solid lines), 5 (model EM2.05, dotted lines) and 0 (model EM2, dashed lines;
no surrounding cloud).
The dash-dot line on the SED graph corresponds to the background SED.}
\label{fig_oem2a}
\centerline{
\includegraphics[width=5.8cm]{3799.f32}
\includegraphics[width=5.8cm]{3799.f33}
\includegraphics[width=5.8cm]{3799.f34}
}
\caption{Intensity profiles
at 90 (a), 170 (b), 450 and 850~$\micron$ (c), for the models
in Fig.~\ref{fig_oem2a} (EM2.20: solid lines, EM2.05: dotted lines, 
EM2: dashed lines).
The horizontal solid lines on the 
profiles correspond to the background intensity at the
wavelength marked on the graph.
In this case (more centrally condensed core than that in Fig.~\ref{fig_oem1b}), the
intensity at the centre of the core at 90~$\micron$ is lower and, thus, the core
is relatively more easily observed in absorption than a less centrally condensed core.
In addition, the increase of the intensity towards the edge of the core at 170~$\micron$,
is larger in this case and thus easier to observe.}
\label{fig_oem2b}
\end{figure*}

In the UV and optical (0.01-1~$\micron$), the radiation coming from the system 
is scattered light and 
direct background radiation (mainly coming from the edge of the cloud 
where the optical depth is small).
In the NIR and MIR (1-50~$\micron$) most of the radiation
 is direct background 
radiation that just passes through the outer, optically thin parts of 
the cloud. This depends on the assumed background radiation field,
the optical depth of the cloud (and hence the dust properties)
and the extent of the cloud.
The  FIR and mm range (60-1300~$\micron$) 
is the most interesting area since the core emits most of its
radiation at these wavelengths. 
Many terrestrial and space-borne observatories cover (or have covered) this range:
{ISOPHOT/{\it ISO} (90, 170 and 200~$\micron$),
SCUBA/JMCT (350-1300~$\micron$), IRAM (1300~$\micron$)
and finally the upcoming {\it SIRTF} 
(3.6-160~$\micron$, to be launched in 2003) 
and {\it Herschel} (75-500~$\micron$, to be launched in 2007).

At 90 microns the core is seen in absorption against the 
background (Figs.~\ref{fig_oem1b} and \ref{fig_oem2b}a). 
The intensity depends on the temperature
exponentially, so
the relatively small increase in temperature towards the
edge of the core can compensate for the rapid decrease
in the density ($\rho\sim r^{-2}$), and the intensity 
increases towards the edge of the core.
For a very centrally-condensed core (e.g. models EM2x, 
Fig.~\ref{fig_oem2b}a)
the decrease towards the centre
is around $\sim 8-10$ MJy${\rm\, sr^{-1}}$ (depending on how deep the core is 
embedded in the cloud; for deeper embedded cores the intensity decrease 
is smaller), and this would be very difficult 
to detect.
For less centrally-condensed cores (e.g. EM1x, Fig.~\ref{fig_oem1b}a)
the decrease is even smaller ($\sim$ 4-6 MJy${\rm\, sr^{-1}}$). 
Thus, very sensitive (say $\sim 1-3 $ MJy${\rm\, sr^{-1}}$) observations are
needed to detect cores in absorption. 
This sensitivity is 
very close to the limits of current instruments, so it is
very difficult to observe embedded prestellar cores at 90~$\micron$.

At wavelengths near the peak of the emission (150-250~$\micron$)
the intensity increases by a small 
amount ($\sim 5-20$ MJy${\rm\, sr^{-1}}$ above the background) 
towards the edge of the cloud and then decreases to
the background intensity (Figs.~\ref{fig_oem1b}b and \ref{fig_oem2b}b).
If the temperature 
increase towards the edge of the core 
is big enough to compensate for the decrease in density,
the outer parts of the core are just visible 
(e.g. models with visual optical depth $\tau_{\rm cloud}=5$). 
On the other hand, if the increase of the temperature is not high enough,
as happens when the core is deeply embedded ($\tau_{\rm cloud}=20$),
then the core cannot be distinguished from the background. 
Thus, our models indicate that cores can be observed at 150-250~$\micron$
only if they are surrounded by a cloud with a relatively small 
visual optical depth $\tau_{\rm cloud}\sim 5$, in which
case the intensity increase is  $\sim 10$ MJy${\rm\, sr^{-1}}$. Cores could
in principle be observed even if they are deeply embedded, provided 
there were accurate observations of $\sim 1$ MJy${\rm\, sr^{-1}}$ at 150-250~$\micron$.
This result agree with the fact that {\it ISO} did not detect 
the prestellar condensations in $\rho$ Oph (Andr\'{e}, Ward-Thompson \& Barsony 2000).

Finally, at submillimetre and millimetre wavelengths (400-1300~$\micron$)
the Rayleigh-Jeans approximation for the Planck function
holds, and the observed intensity is proportional
to the product of the core column density and temperature. Thus, at the edge of
the core the intensity drops considerably because the
temperature increase cannot compensate for the density decrease. 
The core can be easily observed at 400-500~$\micron$, where the contrast
with the background is quite considerable ($\sim 50-150$ MJy${\rm\, sr^{-1}}$).
At wavelengths longer than $\sim$ 600~$\micron$ 
the background radiation becomes important and the core emission is
not much larger than the background emission. For example at 850~$\micron$
(Figs.~\ref{fig_oem1b} and \ref{fig_oem2b}c) the core emission is
only $\sim$ 20-50 MJy${\rm\, sr^{-1}}$ above the background, depending on the density 
profile of the core and how deeply the core is embedded inside the molecular
cloud. 
High accuracy observations are needed 
to observe cores at mm wavelengths, but they are available. For example, 
the sensitivity of IRAM is around $\sim$ 1 MJy${\rm\, sr^{-1}}$.
The peak luminosities at 1300~$\micron$ that we compute with our models 
are comparable with the observed luminosities of Motte et al. (1998).

To check if our results are affected by the extent of the ambient cloud, 
we study a core with the same parameters as the EM2.05
model but embedded in a more extended, less dense cloud. 
In both models the optical depth of the cloud is the same.
As seen in Fig.~\ref{fig_oem_ldens}, the temperature and 
intensity profiles at 90, 170 and 450 $\micron$ 
of the two models are almost identical inside the core.
This result indicates that the only parameter of the ambient cloud 
that is
important in determining the dust temperature and
the SED of a core embedded in 
the centre of a molecular cloud, is
the optical depth of the molecular cloud.

\begin{figure*}
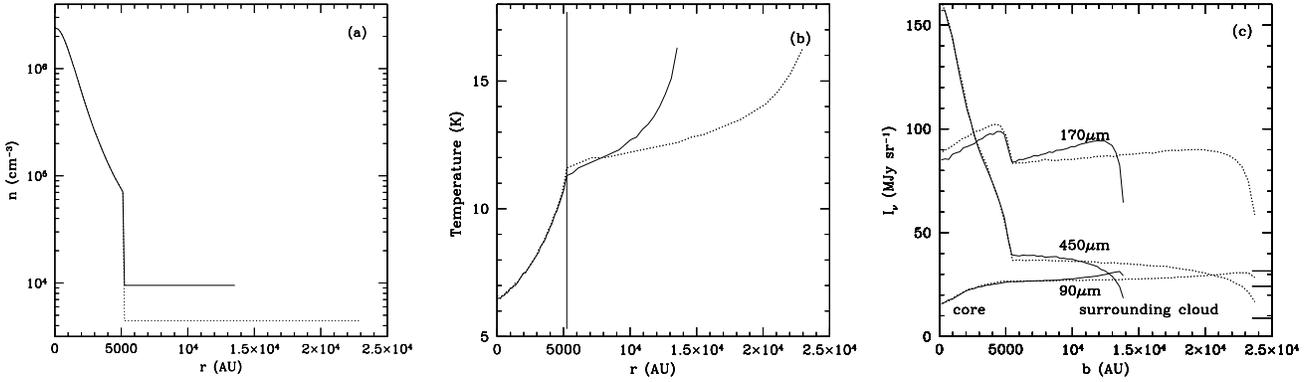
 
\centerline{
\includegraphics[width=5.8cm]{3799.f35}
\includegraphics[width=5.8cm]{3799.f36}
\includegraphics[width=5.8cm]{3799.f37}
}
\caption{Density profiles (a), dust temperature profiles (b) and
intensity profiles (c)
at 90, 170 and 450~$\micron$ 
(the horizontal solid lines on the 
profiles correspond to the background intensity at the
170, 90 and 450 $\micron$, from top to bottom), for a 
supercritical BE sphere at 15~K, with mass 0.8 M$_{\sun}$,
under external pressure P$_{\rm ext}=10^6 {\rm cm}^{-3}$~K, surrounded
by a spherical ambient cloud with visual optical depth 5 and 
 $n_{\rm tot}= 0.96\times 10^4\;{\rm cm}^{-3}$ (model EM2.05, solid lines) and
 $n_{\rm tot}= 0.45 \times 10^4\;{\rm cm}^{-3}$ (same optical depth as before but
more extended cloud, dotted lines). The dust temperature and the intensity
profiles are almost identical inside the core for the two models examined, indicating
that these profiles are determined mainly by the optical depth of the core rather
than its physical extent.}
\label{fig_oem_ldens}
\end{figure*}

\begin{figure*}
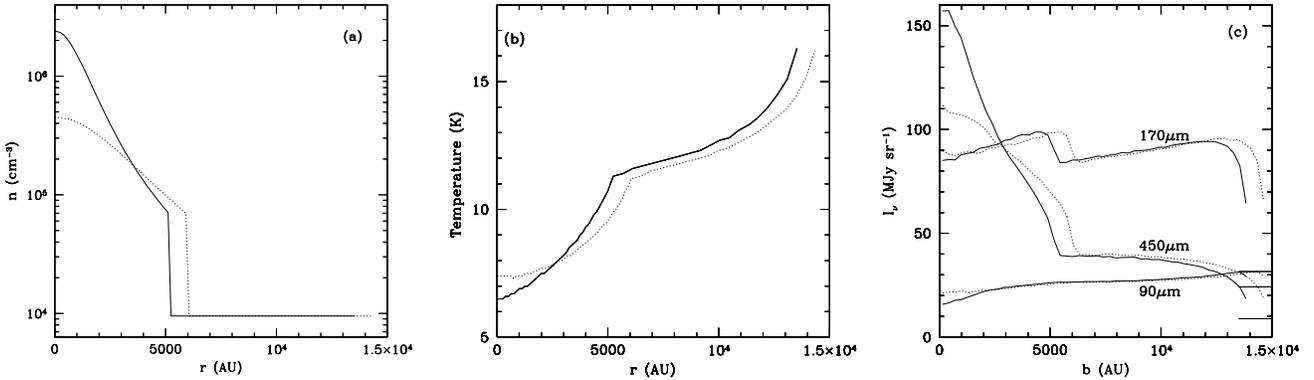
 
\centerline{
\includegraphics[width=5.8cm]{3799.f38}
\includegraphics[width=5.8cm]{3799.f39}
\includegraphics[width=5.8cm]{3799.f40}
}
\caption{Comparison of a subcritical (model EM1.05, dotted lines) 
with a supercritical  (model EM2.05, solid lines) BE sphere, 
embedded in a molecular cloud with visual optical depth 5.
Density profiles (a), dust temperature profiles (b) and
intensity profiles (c)
at 90, 170 and 450~$\micron$.
The horizontal solid lines on the 
profiles correspond to the background intensity at the
170, 90 and 450 $\micron$, from top to bottom.
The dust temperature inside a more centrally condensed core is lower than for
a less centrally condensed core. The core emission is shifted towards
longer wavelengths and, thus,  a supercritical core will emit more radiation at
submm wavelengths than a subcritical core does. }
\label{fig_oem_comp}
\end{figure*}

\subsection{Diagnostics}

In Table~\ref{tab:diagnostics}, we list the peak intensities
(maximum intensity above or below the background)
at wavelengths 90, 170, 450, 850, 1300~$\micron$, 
for cores embedded in molecular 
clouds with visual optical depths 5 and 20. The lower intensity values
correspond to less condensed cores (subcritical) and the higher
intensity values to more condensed cores (supercritical). 
This table indicates that embedded  cores are 
most easily distinguished
from the
background radiation around 450~$\micron$.
The peak emission from 
embedded cores could be as low as $\sim$ 10 MJy${\rm\, sr^{-1}}$ above the background at
1300~$\micron$, but it's at least  $\sim$ 40 MJy${\rm\, sr^{-1}}$ at  450~$\micron$.
The wavelength range between 400-500 seems favourable for observing
embedded cores but the atmospheric transmission is not good in this range 
and space observations are needed. The upcoming
{\it Herschel} space telescope will be operating 
in this range. 

Continuum intensity observations at different submm wavelengths
can be used to determine if a core is subcritical  or supercritical, 
assuming cores can be described as BE  spheres.
If the core is supercritical, it is more condensed in the centre and
the optical depth to the centre of the core is larger than a subcritical,
less condensed core. This means that the dust temperature at the centre of
the core is less for a supercritical core, and the resultant
spectrum is shifted towards longer wavelengths. 
Thus, a supercritical core emits more radiation at longer 
wavelengths (e.g. 450~$\micron$; see Fig.\ref{fig_oem_comp}) than  a
subcritical core. We can exploit the fact that
the intensity at  170-200~$\micron$ varies little for different
cores in the same environment, and
use the colour index 
CI=$m({\rm 450~\micron})- m({\rm 170~\micron})$ to distinguish between
subcritical
and supercritical cores, eliminating this way any uncertainties 
about the distance of the observed cores.
The CI will be larger for supercritical cores.
This result can be used to determine whether a core is subcritical or 
supercritical when the core cannot be resolved 
and the usual density criterion 
($\rho_{\rm centre}/\rho_{\rm edge}>14.1$ for supercritical
spheres) is not useful.

Our models can also be used to estimate the visual extinction of
the ambient cloud surrounding an embedded core.
The outer parts of even deeply embedded cores ($\tau_{\rm cloud}\sim 20-30$)
are expected to be just visible in emission at 170-200~$\micron$ 
($\sim$ 3 MJy${\rm\, sr^{-1}}$ above the background),
whereas cores embedded in a moderate-thick cloud 
($\tau_{\rm cloud}\stackrel{<}{_\sim}5-7$) will be more visible 
($\sim$~10~MJy${\rm\, sr^{-1}}$ above the background), as seen in 
Figs.~\ref{fig_oem1b}b and \ref{fig_oem2b}b. 
The higher the increase in the intensity near the core boundary, the less embedded 
is the core. Thus, very sensitive observations of embedded prestellar cores
at 170-200~$\micron$ ($\sim 1-3$~MJy${\rm\, sr^{-1}}$),
 might allow us to determine the extinction of the cloud
surrounding the core, and thus to estimate roughly the
position of the core inside the molecular cloud. 
However, more sophisticated modelling is required, with 
a more detailed density profile for the cloud and taking into account the
close environment of the core under study.

\begin{table}
\caption{Typical peak$^{\mathrm{*}}$ intensities for embedded cores}
\begin{center}\begin{tabular}{@{}ccc}
\hline
 $\lambda$ ($\micron$) &  \multicolumn{2}{c}{ ${I_\lambda}^{\mathrm{a}}$ (MJy${\rm\, sr^{-1}}$)} \\
 &   $\tau_{\rm cloud}=5$ &  $\tau_{\rm cloud}=20$ \\
\hline\hline
90$^{\mathrm{b}}$   &   5-15  &  $\sim$ 3  \\  
170     &   10-15 &   $\sim$ 3  \\
450     &   55-160      &   40-130  \\ 
850     &   20-80       &   15-70    \\
1300    &   10-40 &   10-25  \\
\hline
\end{tabular}\end{center}
\begin{list}{}{}
\item[$^{\mathrm{*}}$] The term {\it peak} refers to the maximum intensity
above or below the background (as noted)
at a specific wavelength.
\item[$^{\mathrm{a}}$]These are typical approximate peak intensities
for a core embedded in a cloud with visual optical depth $\tau=5$ and
 $\tau=20$.
The deeper the core is embedded the less distinct from
the background is. The lower value corresponds to a subcritical core and the higher
value to a supercritical core.
\item[$^{\mathrm{b}}$] At 90~$\micron$ the core seen in absorption against the 
background.
\end{list}
\label{tab:diagnostics}
\end{table}

\section{Summary}
\label{sec:conclusions}

We have implemented a Monte Carlo radiative transfer method 
to study non-embedded 
and embedded prestellar cores. 
This method discretises the radiation of one or more sources with a 
large number of monochromatic luminosity packets that are injected into 
the system and interact 
stochastically with it. Our code has been tested against
benchmark calculations of other well established radiative transfer codes.
We have also tested our code against the thermodynamic equilibrium test
(the system is illuminated by an isotropic blackbody
radiation field and acquires the same temperature as it) and 
the pure scattering test (only scattering interactions are allowed and
the output intensity is the same as the input one).

We studied cores that are directly exposed to the ISRF and found
similar results (temperature and  intensity profiles) to 
Evans et al. (2001), using a different radiative transfer method.
We extended our study to cores that are embedded inside 
spherical molecular
clouds. We assumed that the ambient cloud has uniform density,
and that the dust composition is the same as that in the embedded core. 
In this case, the radiation incident on the embedded  core is not 
isotropic and cannot be represented by the Black (1994) approximation,
since the ambient cloud shields the core from UV, optical and NIR photons and
enhances the FIR and mm part of the spectrum.
We found that, in this case, the temperature is generally
less than 12~K, even for an ambient cloud with low visual extinction ($\sim$5 mag).
The temperature gradients inside an embedded core
are smaller than in the case of a non-embedded core; deeply
embedded cores are almost isothermal.
Recent studies (Andr\'e et al. 2003) using a different
approach, in which they estimate
the effective radiation field incident on an embedded core from observations, 
also find that the temperature inside embedded cores is
lower than in non-embedded cores.
Previous mass estimates using mm fluxes have assumed 
isothermal cores at temperatures
12-20~K and, consequently, they may have underestimated the masses of the
cores by up to a factor of~2.
However, a more detailed modelling is needed for each specific core for
more accurate mass estimate.

Our models provide a view of cores at a wide range of wavelengths. 
We found that the best wavelength range to observe embedded cores is
 400-500~$\micron$, where the core is easily
 distinguished from the 
background.
Embedded cores could also be observed at 600-1300~$\micron$. 
The contrast of the core radiation 
against the background radiation is not large but very sensitive
observations are available in this range.
At shorter wavelengths the cores are just visible in emission 
(170-200~$\micron$) or in absorption against the background. 
We also found that very sensitive observations at 170-200~$\micron$,
could be used to  estimate the visual extinction of the cloud surrounding
a core, and thus get a rough idea of where the core lies in the
in the environment of the protocluster.
The upcoming {\it Herschel} satellite will be observing in 
the 60-700~$\micron$ range with high sensitivity
and high angular resolution (Andr\'{e} 2002), and will test our models.
Sensitive intensity observations in this range
will also reveal very low-mass condensations present in 
embedded protoclusters, that were previously undetected or 
poorly detected. These observations combined with theoretical models will
enable us to estimate with great accuracy the temperature profile
of resolved prestellar cores. In addition, mm observations from the ground
will provide accurate mass estimates for the cores, and they will constrain
the dust opacity.

Theoretical modelling should be done for each core individually,
taking into account the core
surroundings (ambient cloud, local luminosity sources).
The Monte Carlo approach for the radiative transfer that we have
implemented here is inherently 3D and can treat such asymmetric systems. 
We plan to extend our study to
ellipsoidal cores,  asymmetric ambient clouds and 
anisotropic illuminating radiation fields and, hence, 
model specific cores in embedded protoclusters.

\begin{acknowledgements}

We thank P.  Andr\'{e}, for suggesting the study of embedded cores,
D.~Ward-Thompson and J.~Kirk, for useful discussions on prestellar cores,
M.~Baes for stimulating discussions on Monte Carlo radiative tranfer and
many comments on the paper,
and J. Bouwman for his help in testing the radiative transfer code.
We also thank Z.~Ivezic, for providing his benchmarks calculations for
the code tests,
J.~Black for providing a digital version of 
his estimate of the ISRF and V.~Ossenkopf, for useful comments on
dust opacities in dense cores.
We  acknowledge help from the EC Research Training Network
EC RTN ``The 
Formation and Evolution of Young Stellar Clusters'' (HPRN-CT-2000-00155).

\end{acknowledgements}

\end{document}